\documentclass[a4,11pt]{report}
\usepackage{psfig,subfigure,lscape}
\newcounter{refcount}
\def\thebibliography#1{\section*{References}
 \addcontentsline{toc}{section}{References}
 \list{\arabic{refcount}.}{\usecounter{refcount}%
	\addtocounter{refcount}{0}
         \leftmargin 1.5em
         \itemindent-\leftmargin}
 \def\newblock{\hskip .11em plus .33em minus .07em}
 \sloppy\clubpenalty4000\widowpenalty4000
 \sfcode`\.=1000\relax
}
\begin{document}
\parindent = 1em
\renewcommand{\subfigcapskip}{2pt}
\renewcommand{\subfigcapmargin}{10pt}

\renewcommand{\subcapsize}{\scriptsize}

\newcommand{\goodgap}{%
 \hspace{\subfigtopskip}%
 \hspace{\subfigbottomskip}}

\renewcommand{\thefootnote}{\fnsymbol{footnote}}

\begin{center}
{\Large {\bf The Origin of Cosmic Dust.}}
\end{center}

\vspace{0.5cm}
\large {\bf Loretta Dunne$^{\ast}$, Stephen Eales$^{\ast}$, Rob Ivison$^{\dag}$, Haley Morgan$^{\ast}$,
Mike Edmunds$^{\ast}$}\\
\vspace{0.5cm}
$^{\ast}${\em{Department of Physics \& Astronomy, Cardiff University, 5 The Parade, Cardiff CF24 3YB, UK\/}\\}
$^{\dag}${\em Astronomy Technology Centre, Royal Observatory,
Blackford Hill, Edinburgh EH9 3HJ, UK\/}\\

\normalsize
{\bf Large amounts of dust ($>10^8 \,\rm{M_{\odot}}$) have recently
been discovered in high redshifts quasars$^{1,2}$ and
galaxies$^{3-5}$, corresponding to a time when the Universe was less
than one-tenth of its present age. The stellar winds produced by stars
in the late stages of their evolution (on the asymptotic giant branch
of the Hertzsprung-Russell diagram) are though to be the main source
of dust in galaxies, but they cannot produce that dust on a
short-enough timescale$^6$ ($< 1$ Gyr) to explain the results in the
high-redshift galaxies. Supernova explosions of massive stars (type II) 
are also a potential source, with models predicting 0.2--4
$\rm{M_{\odot}}$ of dust$^{7-10}$. As massive stars evolve 
rapidly, on timescales of a few Myr, these supernovae could be
responsible for the high-redshift dust. Observations$^{11-13}$ of
supernova remnants in the Milky Way, however, have hitherto revealed only 
$10^{-7}-10^{-3}\,\rm{M_{\odot}}$ of dust each, which is insufficient
to explain the high-redshift data. Here we report the
detection of $\sim 2-4 \,\rm{M_{\odot}}$ of cold dust in the
youngest known Galactic remnant, Cassiopeia A. This observation
implies that supernovae are at least as important as stellar winds in
producing dust in our Galaxy and would have been the dominant source
of dust at high redshifts.}

Over the past three decades, many searches for dust in supernova
remnants (SNR) have been made in the mid and far-infrared
(6--100$\mu$m). Remnants must be studied when they are young, before
they have swept up large masses of interstellar material which makes it
difficult to distinguish dust formed in the ejecta from that
present in the ISM prior to the explosion. The handful of Galactic
remnants which are both young and close enough
(Cas A, Kepler and Tycho) have been studied with the Infrared
Astronomical Satellite (IRAS) and the Infrared Space Observatory (ISO), but
although dust at 100--200 K has been detected, the dust mass deduced is
only $10^{-7}-10^{-3}
\rm{M_{\odot}}$, many orders of magnitude lower than the solar mass
quantities predicted.$^{11-13}$ The formation of dust in the recent
supernova 1987A has been implied indirectly from the fading of the
silicate line and increase in 10$\mu$m emission,$^{14}$ although the
quantity is heavily dependant on the assumptions made about the
clumpiness of the dust{\footnote{If the dust is uniform, the
observations imply a very low efficiency of condensation of heavy
elements into dust grains ($\sim 0.001$\%), while if dust is very
clumped this could rise to almost 100\% in some species}. The IRAS/ISO
observations were not sensitive to cold dust at $\leq 25$ K, if the
models are correct about the amount of dust produced in supernovae
(SNe) then the bulk of the dust must be cold. Such dust will emit most
of its radiation at longer wavelengths and so the best place to search
for cold dust emission is at sub-millimetre (submm) wavelengths
(0.3--1 mm). Here, we present our analysis of submm data for
Cassiopeia A, the youngest known Galactic remnant. Cas A is the
brightest radio source in the sky, and still produces significant
synchrotron emission at submm wavelengths. Cas A is believed to be the
remnant of the explosion of a massive (20-30 $\rm{M_{\odot}}$)
progenitor star, which occurred around 320--340 yrs ago at a distance
of 3.4 kpc.$^{15,16}$

Cas A was observed with the SCUBA$^{17}$ array on the JCMT in June
1998. The 850$\mu$m image is presented in Fig.~\ref{850} and shows a
ring-like morphology, similar to the X-ray and radio
images$^{18,19}$. Over two-thirds of the 850$\mu$m emission in the
main ring is synchrotron produced by relativistic electrons spiralling
in the intense magnetic field. Fig.~\ref{radsubsed} presents the
spectral energy distribution (SED) of Cas A from the radio to the
mid-IR, showing the clear excess due to dust at wavelengths of $\leq
850\mu$m. The synchrotron radiation displays a constant power-law
behaviour over more than two decades in frequency with $S_{\nu}
\propto
\nu^{\alpha}$ where $\alpha = -0.72$, meaning that its
contribution to the submm flux can be easily estimated and
subtracted. The 450$\mu$m image is shown in Fig.~\ref{450},
the synchrotron contribution is now only one third and emission from
cold dust dominates which is why the 450 and 850$\mu$m images appear
different. We can remove the synchrotron contribution from the submm
maps by scaling an 83 GHz image,$^{19}$ which is the closest in
frequency and resolution to our images. Fig.~\ref{syncsub850} shows
the result with contours denoting the 450$\mu$m synchrotron subtracted
flux. The general similarity between the two wavelengths, once the
synchrotron has been removed, is strong evidence that this is indeed
emission from cold dust.

The rings on Fig.~\ref{syncsub850} indicate the position of the
forward and reverse shocks as determined from Chandra X-ray
data.$^{18}$ Most of the dust appears to be contained between the two,
where the gas density as traced by the X-rays is greatest. We fitted a
two-temperature grey body SED to the synchrotron corrected IR/submm
fluxes (Fig.~\ref{radsubsed}) determining temperatures of 112 K and 18
K for the two components. The SED parameters and their uncertainties
are given in Table 1. The temperature of the hot dust is easily
explained if it is co-extensive with the X-ray emitting gas ($10^6 -
10^7$ K) and is heated by collisions with fast moving electrons and
ions$^{13}$. The cold dust could be explained in a number of ways and
a more detailed investigation will be presented elsewhere (Dunne et
al. in prep). If the grains are very small ($<0.005\mu m$) they may
cool to 15--20 K in the time between collisions with the gas
particles. However, the dust SED might then be expected to show dust
at all temperatures between 100-20 K, which is not
apparent. Alternatively, the cold dust may be located in a very hot
but diffuse phase of the gas, surrounding the bright, dense X-ray
filaments. If the gas had a density of $<0.01\,
\rm{cm^{-3}}$ then the grains could be this cold. The dust could also be
contained in dense clumps which are in not in pressure equilibrium with
the more diffuse X-ray gas, a model which has been suggested for SN
1987A$^{14}$. Finally, the cold dust could be in thermal equilibrium
with the same X-ray gas which is responsible for heating the hot dust,
if the cold grains are both very large (1-10$\mu$m) and emit more
efficiently than grains in the diffuse ISM. Evidence for unusual
grain properties in Cas A is presented below.

The mass of dust can be estimated from the submm emission using $M_d =
\frac{S_\nu D^2}{\kappa_\nu\, B(\nu, T)}$, where D is the distance to
the remnant (3.4 kpc), $B(\nu, T)$ is the Planck function at
temperature T and $\kappa$ is the dust mass absorption
coefficient. The submm emission is clearly dominated by the cold dust,
therefore the cold temperature is the only one we need be concerned
with. The value of $\kappa$ is the main uncertainty as its value is
not well determined and may vary with physical environment. Dust
masses derived from the SED and different values of $\kappa$ are in
Table 1. Using `standard' $\kappa$ values appropriate to the diffuse
ISM$^{20}$ ($\kappa_3$) gives dust masses which are 
uncomfortably high ($> 7 \,\rm{M_{\odot}}$). Higher $\kappa$ values,
as measured in environments where dust may be newly formed,
amorphous or coagulated (see Table 1) produce more sensible estimates (2 -- 6
$\rm{M_{\odot}}$), thus the dust in Cas A appears to have different
properties to that in the diffuse ISM. It is possible that grain
processing in the diffuse ISM by UV photons and passage through dark
clouds may alter the surface chemistry or shape of grains, causing a
change in their emissivity. At an age of 320 yrs, the dust in Cas A is
relatively pristine and has not yet been subjected to the above
events, this may contribute to its greater emissivity.

Was the dust produced in the supernova explosion, or could it be
pre-existing dust which was swept up by the blast wave? If the
remnant has swept up ISM material with a uniform density$^{19}$ of
0.4--4 H atoms per $\rm {cm^{-3}}$ then only $0.004-0.04
\,\rm{M_{\odot}}$ of dust should be present\footnote{for an ISM
gas-to-dust ratio of 160}. Higher ISM densities are possible, but in
order for the mass of dust inferred in Cas A to have been swept up
densities of the order 200--400 $\rm {cm^{-3}}$ are required. This
would mean Cas A has swept up $\sim 200-400 \,\rm{M_{\odot}}$ of
material which is inconsistent with its dynamical state$^{18,21}$,
believed to be just entering the Sedov stage (i.e. it has swept up
about as much mass as it ejected ($<15 \rm{M_{\odot}}$)). Another
hypothesis for the bright ring seen in the X-ray and radio is that it
is the interaction of the supernova with a stellar wind shell swept up by
the Wolf-Rayet precursor$^{16,22}$. In this case, one could estimate that
somewhere between 4--16 $\rm{M_{\odot}}$ of material could have been
lost in the red-supergiant phase and swept up into the ring. For this
to be the source of the dust we observe requires an implausibly large
quantity of {\em freshly synthesised} elements efficiently condensing
into dust in the stellar wind material$^6$ (which, unlike the supernova ejecta,
is mainly H and He). We conclude that the most likely explanation for
the large mass of dust in Cas A is that it was formed as a result of the supernova
explosion.

Using our result, we estimate the current Galactic dust production
rate from Type-II SNe as $7-18 \,(\times 10^{-3}) \,
\rm{M_{\odot}\,yr^{-1}}$. Stellar wind$^6$ sources produce
$\sim 5 \times 10^{-3}\,\rm{M_{\odot}\,yr^{-1}}$, meaning that SNe are
the dominant source of interstellar dust during most of galactic
evolution. We have shown here that individual SNe are capable of
producing of order of a solar mass of dust in the short times available
in the early universe and therefore the dust observed at high redshift
is likely to have originated in supernovae.

{} 

\section*{Aknowledgments}
We wish to thank Melvyn Wright for kindly providing us with the 83 GHz
image, and Walter Gear and Dave Green for useful discussions. LD is
supported by a PPARC postdoctoral fellowship and HM by a Cardiff
University studentship. We are grateful to Gary Davis for use of
Director's time on the JCMT to obtain the new photometry data.

The authors declare that they have no competing financial interests.

Correspondence and requests for materials should be addressed to
Loretta Dunne\\ (L.Dunne@astro.cf.ac.uk).

\newpage

\begin{table*}
\centering
\caption{\label{table1}\bf Properties of Cas A}
\begin{tabular}{cccccc}
\multicolumn{3}{c}{SED parameters}&\multicolumn{3}{c}{Dust masses
($\rm {M_{\odot}}$)}\\
\multicolumn{1}{c}{$\beta$}&\multicolumn{1}{c}{$\rm{T_{hot}}$
(K)}&\multicolumn{1}{c}{$\rm{T_{cold}}$ (K)}&\multicolumn{1}{c}{$\kappa_1$}&\multicolumn{1}{c}{$\kappa_2$}&\multicolumn{1}{c}{$\kappa_3$}\\
$0.9^{+0.8}_{-0.6}$ & $112^{+11}_{-21}$ & $18.0^{+2.6}_{-4.6}$ & $M_{450} =
2.2^{+1.2}_{-0.2} $ & $M_{450} = 3.8^{+2.0}_{-0.4} $ & $M_{450} = 12.7^{+6.9}_{-1.1} $ \\
 & & & $M_{850} = 2.4^{+0.7}_{-0.3} $ & $M_{850} = 6.2^{+1.7}_{-0.9} $ & $M_{850} = 26.0^{+6.4}_{-4.1} $
\\
\end{tabular}
\flushleft
SED parameters are the best least-squares fit to the data points. The
errors were estimated using a bootstrap procedure in which the
measured fluxes were perturbed according to their Gaussian errors and
re-fitted to find new parameters. The process was repeated 3000 times
and the uncertainties quoted are the 68\% confidence interval derived
from this method. Quoted dust masses are those from the best-fit SED
parameters for each value of $\kappa$, and the ranges reflect the 68\%
confidence interval using the distribution produced by the bootstrap,
essentially the uncertainty in dust mass due to the uncertainty in
fluxes and SED fitting. We have used three values of $\kappa$ which
represent the average from numerous source in the literature for
different grain environments (to be presented in more detail
elsewhere, Dunne et al. in prep). (1) Laboratory measurements and
theoretical modelling suggest that if grains are amorphous or have a
clumpy, aggregate structure $\kappa$ can be very high $0.6-1.1
\,\rm{m^2 kg^{-1}}$ at 850$\mu$m. We take $\kappa_1(450) =
1.5\,\rm{m^2\,kg^{-1}, \,\kappa_1(850) = 0.76 \, m^2
\, kg^{-1}}$.
(2) Environments where dust may be newly formed, or modified by icy
mantles also show high values $0.16-0.8\,
\rm{m^2 kg^{-1}}$. From observations of
evolved stars, VRN (visual reflection nebulae), planetary nebulae and cold clouds in the Galaxy we take
$\kappa_2(450) = 0.88 \, \rm{m^2 \, kg^{-1} \, ,\kappa_2(850) = 0.3 \,
m^2 \, kg^{-1}}$. 
(3) Studies of extragalactic systems and diffuse
ISM dust in the Galaxy support lower values of $\kappa$ in the region
of $0.04-0.15\, \rm{m^2 kg^{-1}}$ at 850$\mu$m. We take $\kappa_3(450) = 0.26 \,\rm{m^2 \,kg^{-1} \, ,\kappa_3(850) = 0.073 \, m^2 \, kg^{-1}}$.
\end{table*}

\newpage

\begin{figure}
\psfig{file=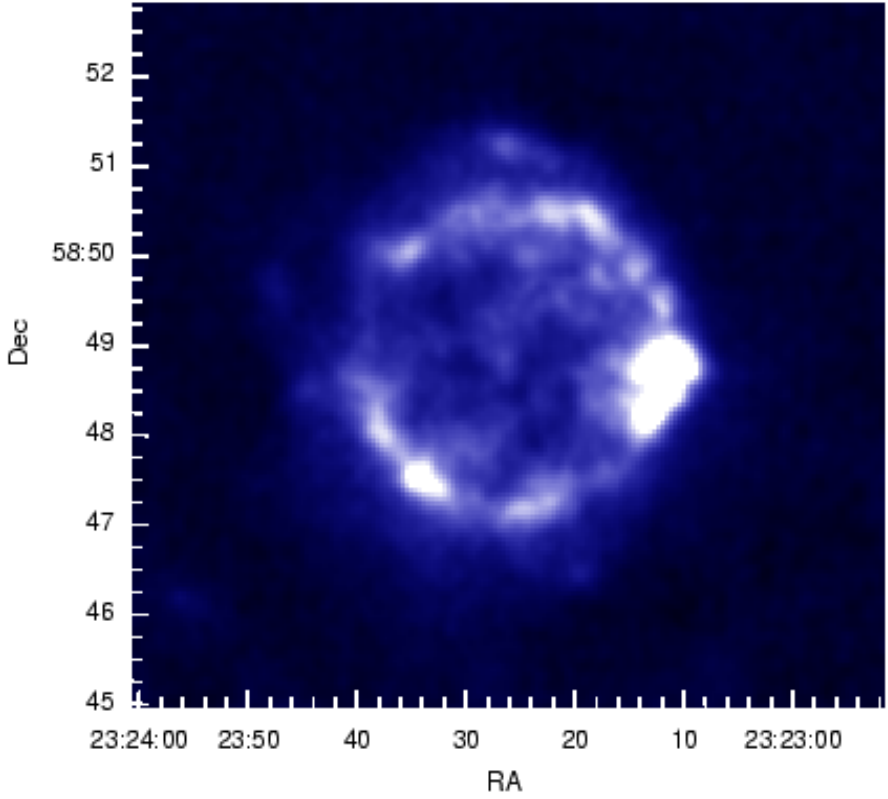,height=12cm,width=12cm,angle=0}
\caption{\label{850} SCUBA 850$\mu$m image of Cas A at a resolution of
15 arcsec. The data were taken in scan map mode and reduced in a
standard way using SURF. Scan map is known to produce artifacts on
large angular scales on the final image and these remain the limitation
on the accuracy of the absolute flux measurement and morphology. On
the 850$\mu$m image, the linear baseline removal left an area of
diffuse emission to the west of the remnant which we believe was an
artifact. After subtraction of a surface which removed this feature
the loss of flux in the remnant was 5 Jy. We confirmed the accuracy of
the absolute flux level of the maps by making new photometric
measurements with SCUBA at several positions on the remnant in
December 2002. The quoted 1$\sigma$ uncertainty on the fluxes is a
combination of calibration errors (5--10\% and 15--20\% at 850/450$\mu$m), plus
the range of fluxes from the maps which agreed with the
photometry. The integrated fluxes in were corrected for the
error lobe contribution (derived from maps of Uranus) by dividing by 1.2 at 850$\mu$m and 2.0 at
450$\mu$m.}
\end{figure}

\newpage

\begin{figure}
\psfig{file=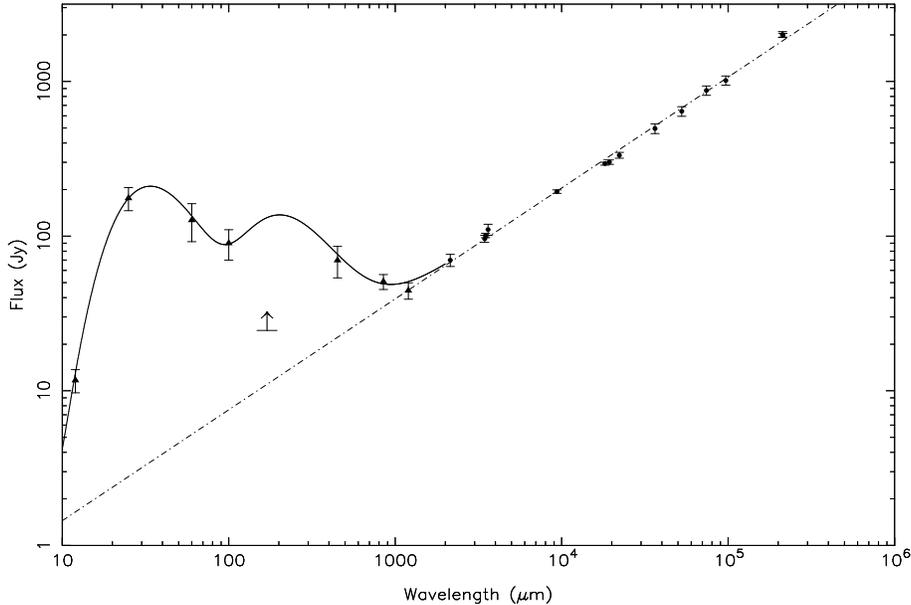,height=8cm,width=12cm,angle=-90}
\caption{\label{radsubsed} The SED of Cas A from the mid-IR to the
radio. All fluxes are integrated. The radio spectrum is fitted with a
power law slope with spectral index $\alpha = -0.72$. The radio
data$^{19,23-28}$ have been scaled to the epoch of our Cas A
observation (1998.4) using a recent study$^{24}$ which found that the
flux decrease of Cas A was independent of frequency at $\approx
0.6-0.7 \% \rm yr^{-1}$. The IRAS points are the averages of the
literature values$^{13,29}$ with the error bars indicating the range
of fluxes. SCUBA fluxes are $50.8 \pm 5.6$ Jy at 850$\mu$m and $69.8
\pm 16.1$ Jy at 450$\mu$m. The synchrotron contribution is 34.9 Jy at
850$\mu$m and 22.1 Jy at 450$\mu$m. The fit to the FIR/submm points is
for a 2 temperature grey-body with $\beta=0.9$, $T_w = 112$ K,
$T_c=18$ K and for 700 times more mass in the cold component than in
the hot. We have fitted 5 parameters to 6 data points and therefore
the fit is just constrained. The uncertainties in the SED parameters
are given in Table 1. Clearly, the submm points at 450 and
850$\mu$m lie above the extrapolation of the radio synchrotron
spectrum. The point at 1.2mm (ref. 28) is not of sufficient accuracy to
determine an excess, but is consistent with our higher frequency
measurements. Longer wavelength ISO measurements$^{30}$ of Cas A at
170$\mu$m suggested that dust at $\sim 30$ K may have been present
(implying of order 0.15 $M_{\odot}$ of dust), however the poor angular
resolution of ISO at these wavelengths made a separation of the
remnant and background emission impossible. The ISO flux is therefore
a lower limit (R. J. Tuffs, private communication).}
\end{figure}

\newpage

\begin{figure}
\psfig{file=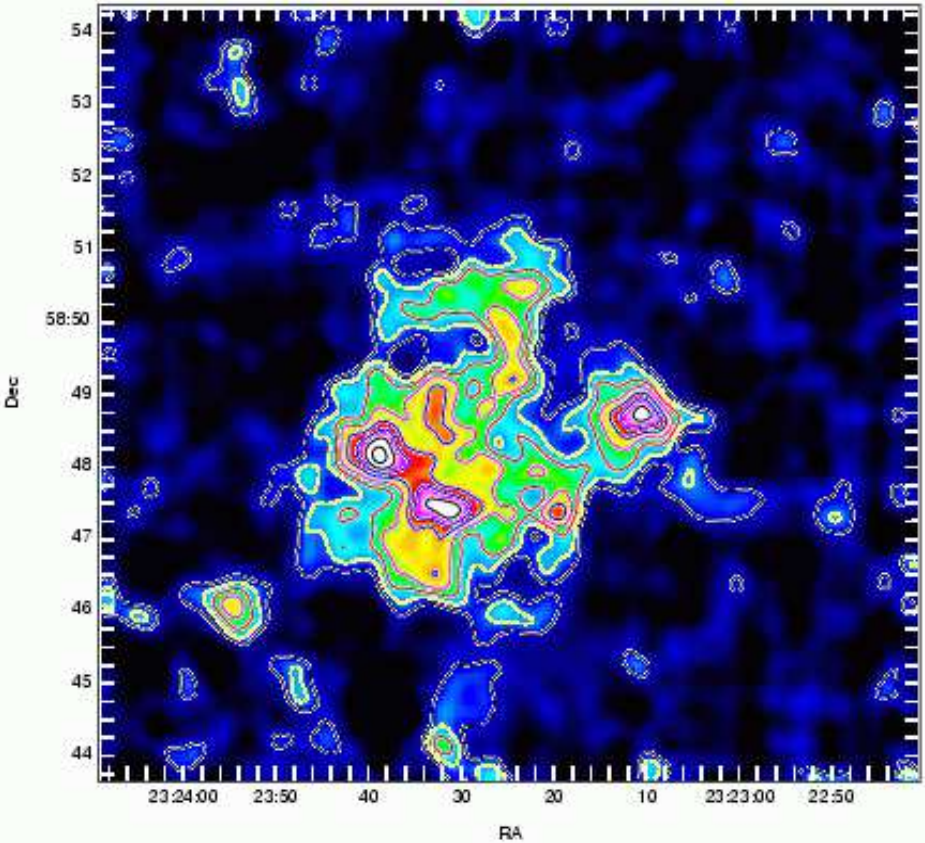,height=12cm,width=13cm,angle=0}
\caption{\label{450} SCUBA 450$\mu$m
map, smoothed with a 21$^{\prime\prime}$ gaussian. Colours and
contours both represent the 450$\mu$m emission. Contours start at
2$\sigma$, with intervals of 1 $\sigma$. The difference in morphology
between this image and that at 850$\mu$m (Fig.~\ref{850}) is because
emission from cold dust dominates at 450$\mu$m, while two-thirds of
the emission at 850$\mu$m is synchrotron.}
\end{figure}

\newpage

\begin{figure}
\psfig{file=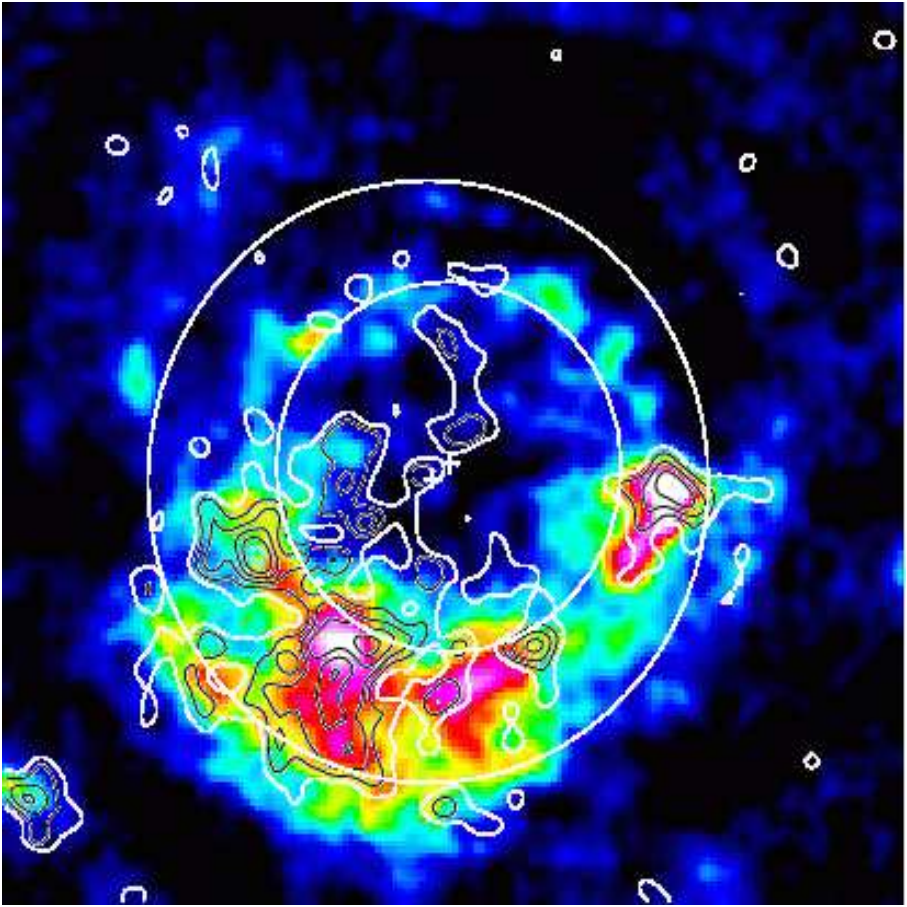,height=12cm,width=12cm,angle=0}
\caption{\label{syncsub850} The 850$\mu$m emission once the
synchrotron has been subtracted using an 83 GHz image$^{19}$. The box
is 8.4 arcmin by 7.8 arcmin, north is up and east is left. Colours
represent the 850$\mu$m intensity, contours are the 450$\mu$m emission
with the synchrotron subtracted, starting at 3$\sigma$ with increments
of $+1\sigma$. The rings and
crosses indicate the location (and centroids) of the forward and
reverse shocks as determined from Chandra X-ray data$^{18}$. The
forward shock is at a mean radius of $153\pm 12^{\prime\prime}$ and
the reverse shock is at a mean radius of $95\pm
10^{\prime\prime}$. The bulk of the dust emission appears to be
bounded by the shocks, where the gas density is highest. There is a
noticeable asymmetry in the dust emission. }
\end{figure}






\end{document}